\documentclass[conference]{IEEEtran}
\usepackage{times}
\usepackage{cite}
\usepackage{algorithmic}
\usepackage[pdftex]{graphicx}
\usepackage{amsmath}
\usepackage{amsfonts}
\usepackage{amsthm,amssymb}
\usepackage{fixltx2e}
\usepackage{url}
\usepackage{booktabs}
\usepackage{flushend}

\newtheorem{definition}{Definition}

\newcommand{\sign}[1]{\textsuperscript{#1}}

\begin{document}
%
\title{Speeding up SOR Solvers for Constraint-based GUIs with a Warm-Start Strategy}

\author{%
{Noreen Jamil, Johannes M\"uller, Christof Lutteroth and Gerald Weber}%
\vspace{1.6mm}\\
\fontsize{10}{10}\selectfont\itshape
Department of Computer Science\\
University of Auckland\\
Private Bag 92019, Auckland, New Zealand\\
\fontsize{9}{9}\selectfont\ttfamily\upshape
\{njam031, jmue933\}@aucklanduni.ac.nz, \{lutteroth, gerald\}@cs.auckland.ac.nz\\
}
\maketitle
\begin{abstract}
Many computer programs have graphical user interfaces (GUIs), which need good layout to make efficient use of the available screen real estate.
Most GUIs do not have a fixed layout, but are resizable and able to adapt themselves.
Constraints are a powerful tool for specifying adaptable GUI layouts:
they are used to specify a layout in a general form, and a constraint solver is used to find a satisfying concrete layout, e.g.\ for a specific GUI size.
The constraint solver has to calculate a new layout every time a GUI is resized or changed, so it needs to be efficient to ensure a good user experience.
One approach for constraint solvers is based on the Gauss-Seidel algorithm and successive over-relaxation (SOR).

Our observation is that a solution after resizing or changing is similar in structure to a previous solution.
Thus, our hypothesis is that we can increase the computational performance of an SOR-based constraint solver if we reuse the solution of a previous layout to warm-start the solving of a new layout.
In this paper we report on experiments to test this hypothesis experimentally for three common use cases:
big-step resizing, small-step resizing and constraint change.
In our experiments, we measured the solving time for randomly generated GUI layout specifications of various sizes.
For all three cases we found that the performance is improved if an existing solution is used as a starting solution for a new layout.
\end{abstract}
\vspace{1\baselineskip}

\section{Introduction}

Various numerical methods have been introduced to solve linear problems as they appear in engineering, mathematics and computer science.
These methods can be divided into direct and iterative methods.
Direct methods aim to calculate an exact solution in a finite number of operations, whereas iterative methods begin with an initial approximation and usually produce improved approximations in a theoretically infinite sequence whose limit is the exact solution~\cite{Saad:2003_iterative_methods}.

Many linear problems are sparse, i.e.\ most linear coefficients in the corresponding matrix are zero so that the number of non-zero coefficients is \emph{$O(n)$} with $n$ being the number of variables~\cite{Kunis:Sparse-Trigonometric-Polynomials}.
Iterative methods do not spend processing time on coefficients that are zero.
Direct methods, in contrast, usually lead to fill-in, i.e.\ coefficients change from an initial zero to a non-zero value during the execution of the algorithm.
In these methods we therefore may weaken the sparsity property and may have to deal with more coefficients, which makes the processing time slower.
Therefore, iterative methods are often faster than naive direct methods in such cases.
In this paper, we are concerned with a domain where sparse problems occur frequently, namely constraint-based graphical user interface (GUI) layout.
Constraint-based layout models have been studied for quite a while in the research community~\cite{Stuckey:arithmetic-constraints,
Weber:High-Level-Constraints,scoditti2009new,ZeidlerCHINZ2012} and attracted attention recently because of a newly introduced layout model in the Cocoa API of Apple's Mac~OS~X\footnote{Cocoa Auto Layout Guide, 2012, \url{http://developer.apple.com}}.

A common iterative method is successive over-relaxation (SOR)~\cite{David:Iterative}.
Starting with an initial guess, it repeatedly iterates through the constraints of a linear specification, refining the solution until a sufficient precision is reached.
For each constraint it chooses a \emph{pivot variable}, and changes the value of that variable so that the constraint is satisfied.
In order to use SOR for GUI layout, certain considerations and extensions are necessary.

First, linear problems in GUI layout are often over-determined and have many inequality constraints, leading to non-square coefficient matrices.
This leads to the problem of \emph{pivot assignment}: this is the problem of choosing a pivot variable for each constraint so that the iterative method converges.
The standard SOR algorithms choose for each constraint the pivot variable on the diagonal of the problem matrix.
In over-determined systems we do not have a main diagonal, therefore we make use of the pivot assignment algorithms proposed in~\cite{Jamil:Linear-Relaxation}.

Second, constraint-based GUI specifications may contain conflicting constraints, rendering a specification infeasible.
To deal with conflicts, \emph{soft constraints} need to be supported.
In contrast to the usual \emph{hard constraints}, which cannot be violated, soft constraints may be violated as much as necessary if no other solution can be found.
Soft constraints can be prioritized so that in a conflict between two soft constraints only the soft constraint with the lower priority is violated~\cite{Wilson:Constraint-hierarchies}.
Using only soft constraints has the advantage that a problem is always solvable, which cannot be guaranteed if hard constraints are used.
In this work, we use the soft constraint algorithms proposed in~\cite{Jamil:Linear-Relaxation}.

A GUI layout specification has to be solved whenever the conditions under which a GUI is displayed or the GUI itself change.
Most GUIs can be resized, e.g.\ to adapt to different screen sizes or to let the user choose an appropriate size dynamically.
Sometimes GUIs need to be changed dynamically to adjust to content of different sizes.
Each time a GUI is resized or changed, the existing GUI layout specification is changed and a new specification is created.
However, the new specification is similar to the previous one because the widgets and their relations typically stay the same.
Usually, only some size parameters change.
For example, Figure~\ref{fig:resizing} shows a GUI that is resized, with the corresponding constraint specifications.
Only the height constraint at the beginning of the specification is changed.

As a consequence, constraint solvers for GUI layout usually have to solve specifications that are similar to the specification that has been solved previously.
For that reason, it seems plausible that the previous solution is a good initial value for the iterative solving process -- something that is known as a warm-start strategy.
This leads us to the following hypothesis:
\emph{Using a previous solution of a GUI constraint specification to warm-start an SOR solver reduces the solving time.}

We tested the hypothesis by considering three common use cases where GUIs are changed during runtime.
First, small-step resizing, where the GUI size is changed by a small amount, e.g.\ when it is resized by a user dynamically.
Second, big-step resizing, where the GUI size is changed by a larger amount, e.g.\ when the GUI size is maximized.
And third, changes of several constraints, e.g.\ when the sizes of labels are adjusted for a different language.
The solving time when using SOR with and without a warm-start strategy was compared for the three use cases, using randomly generated layout specifications of different sizes.

Section~\ref{sec:relatedwork} sums up related work about constraint solving and warm-start strategies.
Section~\ref{sec:background} puts this research into context with background information about constraint-based GUI-layout, SOR, pivot assignment and conflict resolution.
Section~\ref{sec:experiment} presents the methodology and results of our performance experiment.
Section~\ref{sec:conclusion} concludes the paper.

\section{Related Work}\label{sec:relatedwork}

The overall problem, solving linear systems for constraint-based GUIs, is related to solution procedures for over-determined linear systems in general and constraint-based GUIs in particular.
Several direct and iterative methods exist, which can solve over-determined systems in a least-square sense~\cite{bjorck1996numerical}.
Examples are QR-factorization~\cite{bjorck1996numerical}, the simplex algorithm~\cite{Dantzig:Linear-Programming}, the conjugate gradient method~\cite{Golub:Matrix-Computation} and the GMRES-method~\cite{Golub:Matrix-Computation}.
They are the basis for solvers specifically designed to solve problems of constraint-based GUIs.
Some are based on direct methods, for example HiRise and HiRise2~\cite{Hosobe:Simplex-Based} but the vast majority of existing solvers is based on convex optimization approaches and uses slack variables and an objective function~\cite{Badros:cassowary,Stuckey:arithmetic-constraints,Weber:High-Level-Constraints}.
These methods can handle simultaneous constraints, i.e.\ constraints that depend on each other.
In that respect they are superior to local propagation algorithms, such as DeltaBlue~\cite{Freeman:Delta-Blue} and SkyBlue~\cite{Skyblue:local-propagation-constraint-solver}, which cannot do so.

Approaches related to warm-start strategies have been proposed in numerous previous works.
Lessard~\cite{Lessard:08} analyzed computational speed and the effect of warm-starting for different iterative methods with large systems, including the multigrid method, the preconditioned conjugate-gradient method, and several new variants of these methods.
Using a previous estimate for initializing an iterative scheme could reduce computation time significantly.
Wright et al.~\cite{Wright_et_al_2009} used a warm-start strategy to speed up gradient projection for sparse reconstruction (GPSR) and iterative shrinkage/thresholding (IST) algorithms.
Other methods for accelerating convergence by using warm-start techniques in iterative solution procedures are described in~\cite{Ren:Warm-Start,Forsgen2006,Fletcher:1987:Active}.
The use of warm-start strategies for constraint-based GUI layout problems using SOR has not been explored before.

\section{Background}\label{sec:background}

To put our study into context, this section gives a short overview of constraint-based GUIs, SOR, and the extensions of SOR necessary to solve GUI layout specifications.

\subsection{User Interface Layout as a Linear Problem} \label{UILayout}

Constraints are a suitable mechanism for specifying the relationships among objects.
They are used in the area of logic programming, artificial intelligence and GUI specification.
They can be used to describe problems that are difficult to solve, conveniently decoupling the description of the problems from their solution.
Due to this property, constraints are a suitable way of specifying GUI layouts, where the objects are widgets and the relationships between them are spatial relationships such as alignment and proportions.
In addition to the relationships to other widgets, each widget has its own set of constraints describing properties such as minimum, maximum and preferred size.

\begin{figure}[tb]
\begin{center}
\includegraphics[width=\columnwidth]{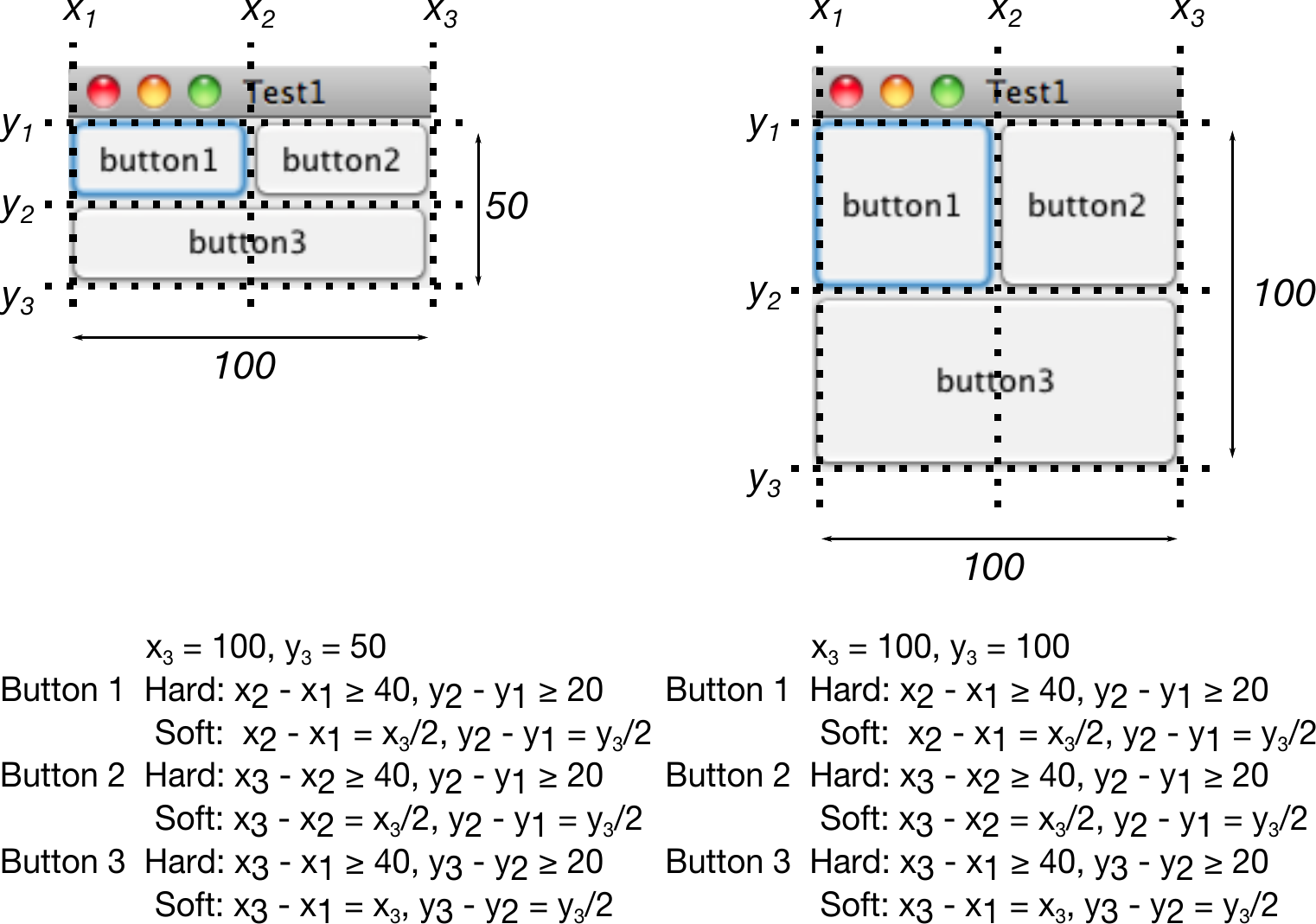}
\caption{A GUI constraint specification before and after resizing} \label{fig:resizing}
\end{center}
\end{figure}

GUI layouts are often specified with linear constraints~\cite{Weber:High-Level-Constraints}.
The positions and sizes of the widgets in a layout translate to variables.
Constraints about alignment and proportions translate to linear equations, and constraints about minimum and maximum sizes translate to linear inequalities.
There are constraints for each widget that relate each of its four boundaries to another part of the layout, or specify boundary values for the widget's size, as shown in Figure~\ref{fig:resizing}.
As a result, the direct interaction between constraints is limited by the topology of a layout, resulting in sparsity of the linear specification.

During application runtime, GUIs need to be adapted to changing conditions such as the available GUI size.
This is done by changing some of the constraints, typically a small number.
For example, the overall size of the GUI is typically specified in by constraints: one for the width and one for the height.
When the GUI size changes, only these two constraints need to be adjusted, as shown in Figure~\ref{fig:resizing}.
Another typical situation where constraints need to be changed is when preferred sizes change.
For example, if the language settings are changed in an application, the preferred sizes of textual labels have to adjust to the new language.

\subsection{Successive Over-Relaxation (SOR)} \label{LinearRelaxation}

Most of the research on iterative methods deals with iterative methods for solving linear systems of equalities and inequalities for sparse square matrices, the most important method being SOR, also known as linear relaxation.
This section summarizes the most important findings.

The best-known iterative method for solving linear constraints is the Gauss-Seidel method~\cite{Bhatti:Numerical-Analysis}.
Given a system of \emph{n} equations and \emph{n} variables of the form
\begin{equation}\label{linearequation}
    Ax=b
\end{equation}
we can rewrite the equation for the \(i\)th term as follows
\begin{equation}\label{linearequationforithterm}
x_{i}=\frac{1}{a_{ii}}({b_{i}-\sum_{j=1}^n
    a_{ij}{x_{j}}} ).\\
\end{equation}

The variable $x_i$, which is brought onto the left side, is called the \emph{pivot variable}, and $a_{ii}$ is the \emph{pivot coefficient} or \emph{pivot element} chosen for row $i$.
An initial estimate for $x$ is chosen, which usually does not fulfill the equations.
The algorithm refines the estimate by repeatedly replacing all individual $x_{i}$ so that the $i$th eqation becomes fulfilled.
This is done in round-robin fashion, and one full run through all $n$ equations is one  iteration, $r$ being the iteration number.
We can therefore write the process as
\begin{equation}
x_{i}^{r+1}=\frac{1}{a_{ii}}({b_{i}-\sum_{j=1}^{i-1}
    a_{ij}{x_{j}^{r+1}}-\sum_{j=i+1}^{n}
    a_{ij}{x_{j}^{r}}} ).\\
\end{equation}
The algorithm iterates until the relative approximate error is less than a pre-specified tolerance.

SOR, also known as linear relaxation, is an improvement of the Gauss-Seidel method~\cite{David:Iterative}.
It is used to speed up the convergence of the Gauss-Seidel method by introducing a parameter $\omega$, known as relaxation parameter, so that
\begin{align}\label{Linearrelaxationmethodformula}
x_{i}^{r+1}=\frac{\omega}{a_{ii}}({b_{i}-\sum_{j=1}^{i-1}
    a_{ij}{x_{j}^{r+1}-\sum_{j=i+1}^{n}a_{ij}x_{j}^{r}}})
   +(1-w)x_{i}^{r}.
\end{align}
This reduces to the Gauss-Seidel method if $\omega=1$.
It is known as over-relaxation if \emph{$\omega>1$}, and known as under-relaxation if \emph{$\omega<1$}.

\begin{definition}
The spectral radius of a matrix is the maximum of the absolute values of its eigenvalues.
\end{definition}
The convergence rate of the SOR method depends on the spectral radius of the coefficient matrix of the problem.
The smaller the spectral radius is, the faster the SOR method converges.
We usually have well-conditioned coefficient matrices in the GUI layout domain for which the SOR method converges.

SOR supports linear equalities as well as inequalities.
Inequalities are handled similar to equalities~\cite{Agmon:Relaxation-Method,Motzkin:Relaxation-Method}:
in each iteration, inequalities are ignored if they are satisfied, and otherwise treated as if they were equalities.

\subsection{Pivot Assignment}

In the case of square coefficient matrices, pivots are selected on the main diagonal.
This is not possible in the over-determined case.
Since the diagonal elements do not lend themselves naturally as pivot elements if the matrix is non-square, we need to explicitly select a pivot element for each constraint.
In other words, we need to determine a \emph{pivot assignment}.

\begin{definition}
A pivot assignment is an assignment of constraints to variables.
\[\gamma\colon \mbox{Constraints} \rightarrow \mbox{Variables}\]
\end{definition}

In previous work, we suggested algorithms which select a pivot element randomly or according to some criteria~\cite{Jamil:Linear-Relaxation}.
We use our random pivot assignment algorithm for the study presented in this paper.
This algorithm assigns the pivot variable for each constraint randomly in each iteration, so that the assignment varies over the iterations.

It is not inherently obvious that randomized assignments work for the linear relaxation approach, but it is the simplest approach that may work.
Although the random algorithm does generally not make the optimal assignment with regard to convergence, it reduces the effect of bad assignments while allowing for good assignments.
In particular, it is guaranteed that every suitable variable will be chosen as pivot variable at some point.
The general assumption underlying randomized algorithms is that the effect of good choices outweighs the effect of bad choices.

In the general case constraint-based GUIs are over-determined, which can result in conflicts between constraints of the problem.
A proper pivot assignment algorithm alone is not sufficient to deal with such cases.
A technique to handle conflicts between constraints, e.g.\ in the form of soft constraints, is required.

\subsection{Conflict Resolution}\label{SoftConstraints}

To resolve conflicts in over-determined systems, soft constraints are introduced.
A natural way to support soft constraints is to assign priorities to all constraints.
These priorities can be defined as a total order on all constraints that specifies which one of two constraints should be violated in case of a conflict.

In our study we use the constraint insertion algorithm proposed in previous work~\cite{Jamil:Linear-Relaxation}.
This algorithm tests constraints incrementally.
We start with an empty set $E$ of enabled constraints.
Iterating through the constraints in order of descending priority, we add each constraint tentatively to $E$ (``enabling'' it), and try to solve the resulting specification.
If a solution is found, then we proceed to the next constraint.
If no solution is found within a fixed maximum number of iterations, then the tentatively added constraint is removed again.
In that case, the previous solution is restored and we proceed to the next constraint.

\section{Experiment}\label{sec:experiment}

In this section, we evaluate the use of warm-starting for GUI layout problems using SOR.
We tested specifically the effect of warm-starting a constraint solver on the performance in terms of computation time.

\subsection{Methodology}

We conducted the experiments with our implementation of an SOR solver for GUI layout, which uses random pivot assignment and constraint insertion as a conflict resolution strategy.
We used two versions of that solver:
the first version started every solver run with an initial solution \(x = (0, \ldots, 0)\),
the second version started every solver run with the optimal solution from the previous run \(x = x^*_{\text{prev}}\).

We evaluated the following three use cases:
\begin{enumerate}
	\item {\em Small-step resizing}: The width and height of the window was randomly changed by a value in between 0 and 3 pixels.
	\item {\em Big-step resizing}: The width and height of the GUI window was randomly changed by a value between 4 and 3000 pixels.
	\item {\em Constraint change}: 10 per cent of all constraints of a GUI were randomly changed.
\end{enumerate}
Small-step resizing occurs in practice when a window is continuously resized by dragging a window border.
Big-step resizing occurs when a GUI is initially loaded on different screens, when a GUI is switched to or from full-screen mode, or when the orientation of a screen is changed.
Constraint changes as in use case 3 occur, for example, when several preferred sizes change as a result of changing the language of an application.

Layout specifications were randomly generated using the parameterized algorithm described in~\cite{Weber:High-Level-Constraints}.
The problem size was varied from 0 to 201 areas.
For each area 4 constraints are added, which specify the position of the area in the layout.
Additionally, a specification needs 4 constraints to define the size of the window.
So we started with a problem of 4 constraints and ended with a problem of 808 constraints.
For each size, 10 random layouts were evaluated.
For each of the three use cases, each of these random layouts was changed 20 times, and the solving time was measured.
A linear relaxation parameter of 0.7 and a tolerance of 0.01 were used for solving.
The measurements were performed on a desktop computer with Intel i5 3.3GHz processor and 64-bit Windows 7 running an Oracle Java virtual machine.

\begin{table}[htb]
\begin{center}
\begin{tabular}{rl}
  \toprule
 Symbol & Explanation \\
  \midrule
 \(\beta_0\) & Intercept of the regression model\\
 \(\beta_{1-3}\) & Estimated model parameters\\
 \(c\) & Number of constraints\\
  \(T\) & Measured time in milliseconds\\
   \(R^2\) & Coefficient of determination\\
\bottomrule
\end{tabular}
\caption{Symbols}
\label{tab:symbs}
\end{center}
\end{table}

To compare the performance of both versions of the solver we used a regression model
\[
T = f(c) + \epsilon
\]
and examined the estimated model visually and numerically.
See Table~\ref{tab:symbs} for an explanation of the symbols used.

\subsection{Results}

\begin{figure}[tb]
	\centering
		\includegraphics[width=1.00\columnwidth]{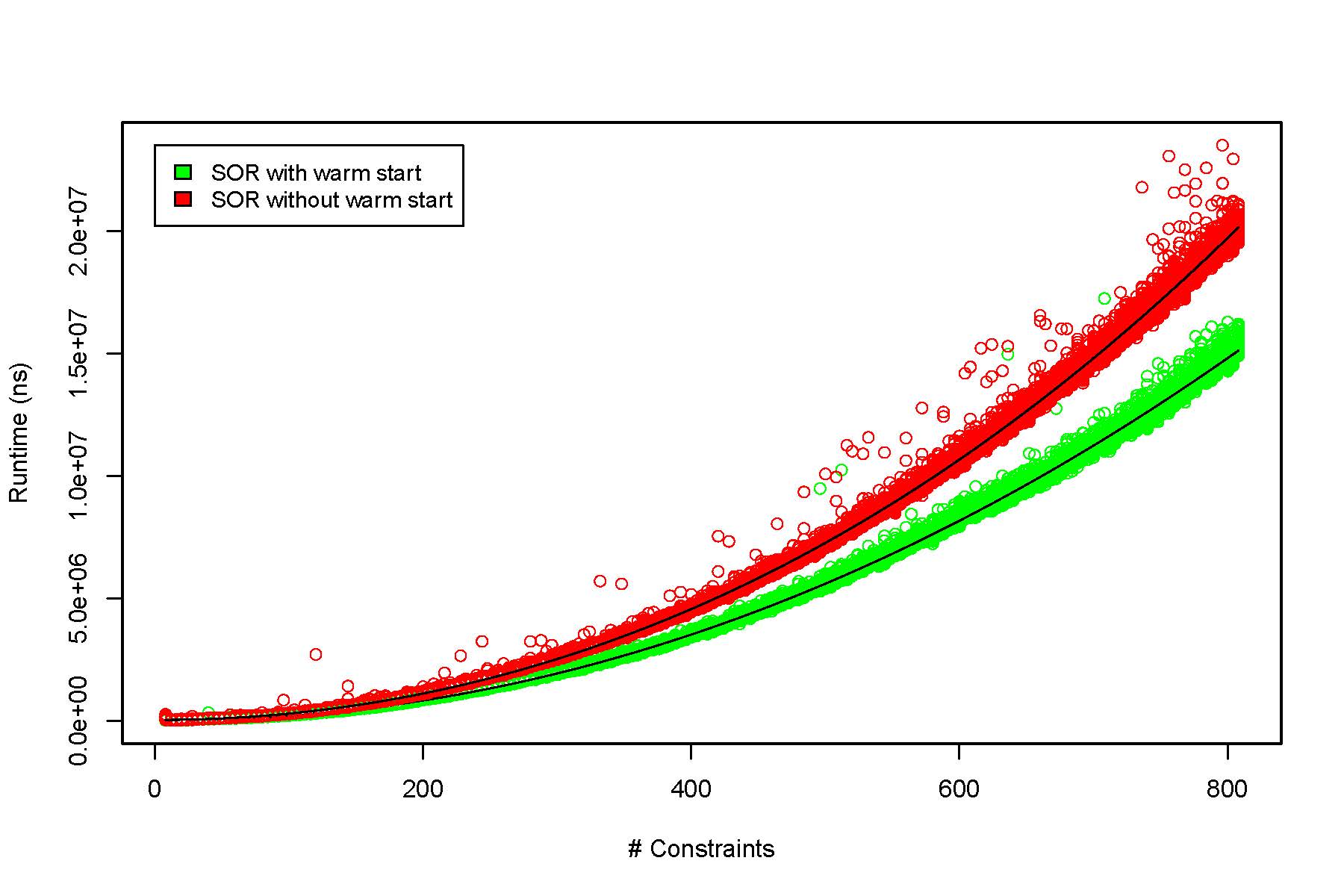}
	\caption{Small-step resizing performance results}
	\label{fig:smallresizing}
\end{figure}

\begin{figure}[tb]
	\centering
		\includegraphics[width=1.00\columnwidth]{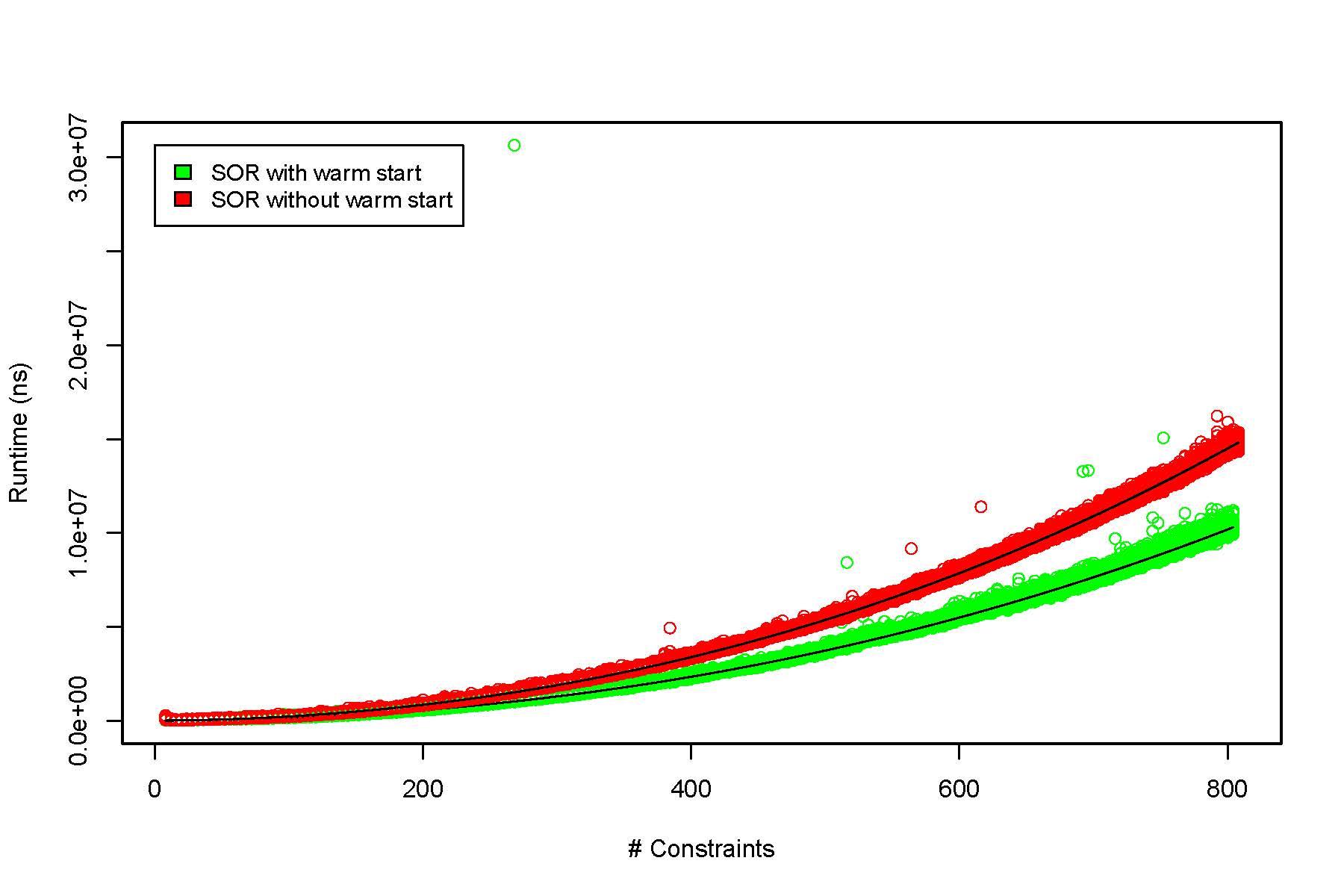}
	\caption{Big-step resizing performance results}
	\label{fig:bigresizing}
\end{figure}

\begin{figure}[tb]
	\centering
		\includegraphics[width=1.00\columnwidth]{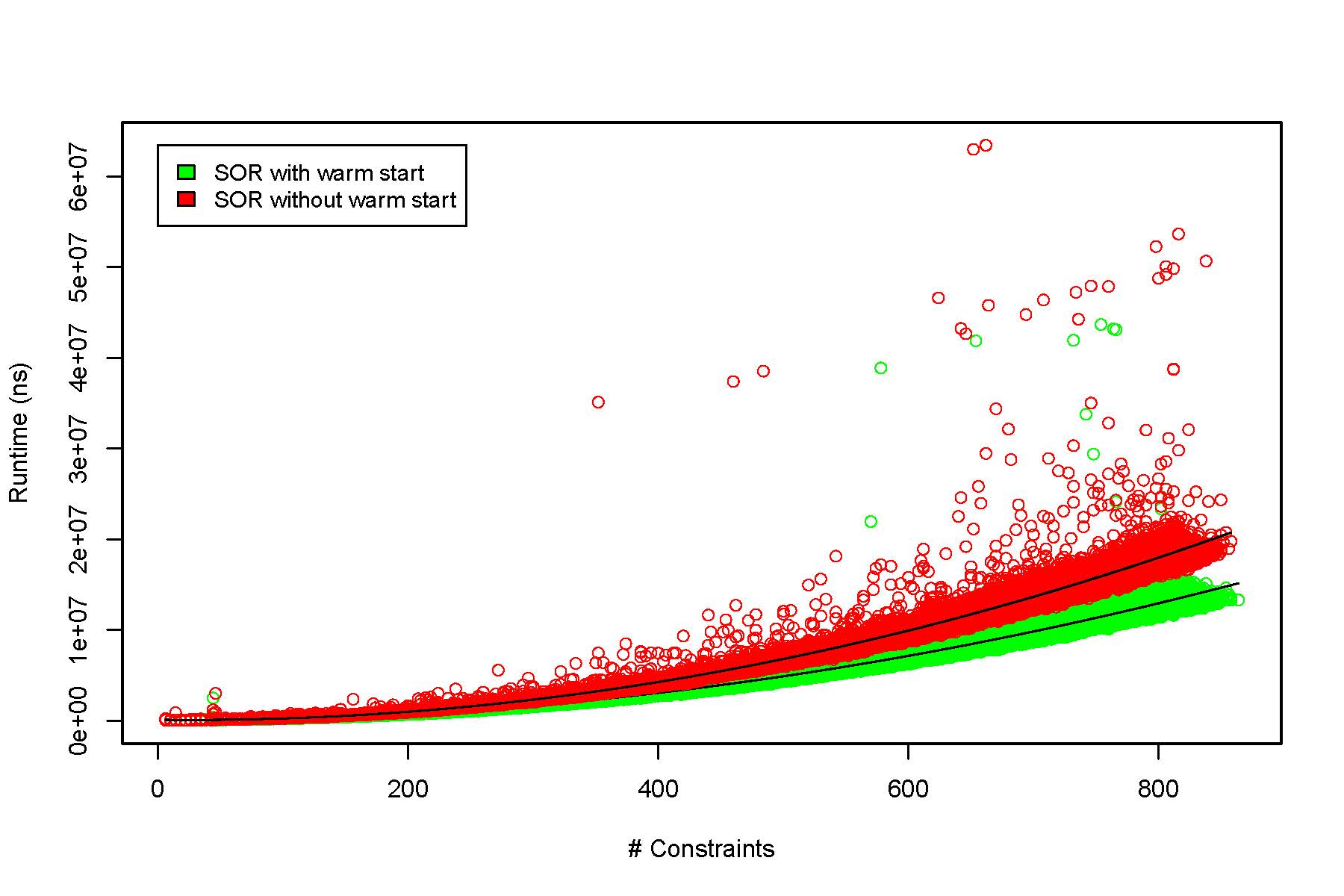}
	\caption{Constraint change performance results}
	\label{fig:constraintchange}
\end{figure}

To identify the performance trend of the solvers, we tried different regression functions $f$ (linear, quadratic, log, cubic).
We found that the best fitting model is the polynomial model
\[
T = \beta_0 + \beta_1c + \beta_2c^2 + \beta_3c^3 + \epsilon.
\]
Key parameters of the regression models are depicted in Table~\ref{tab:reg1}.
A graphical representation of the measurements and the models can be found in Figures~\ref{fig:smallresizing}, \ref{fig:bigresizing} and~\ref{fig:constraintchange}.
The results suggest a better performance of the solver with the warm-start strategy for all three use cases.

\begin{table*}[tb]\begin{center}
\caption{Regression models for solvers with and without warm-start strategy}
\label{tab:reg1}
\begin{tabular}{rlllll}
  \toprule
\multicolumn{1}{c}{\bf Strategy}                       &\multicolumn{1}{c}{\(\beta_0\)} &\multicolumn{1}{c}{\(\beta_1\)} & \multicolumn{1}{c}{\(\beta_2\)}  & \multicolumn{1}{c}{\(\beta_3\)} & \(R^2\)  \\
  \midrule
Small-step resizing with warm start & \( 6.508\cdot 10^{4}\)\sign{***}& \( -8.940\cdot 10^{2}\)\sign{***} & \(23.55\)\sign{***} & \(\hspace{2.3mm} 7.576\cdot 10^{-4}\)\sign{***}& \(0.999\)\\
Small-step resizing without warm start      &\(4.104\cdot 10^{4}\)\sign{***} & \(\hspace{2.3mm} 68.00\)\sign{***} & \(26.05\)\sign{***} & \(\hspace{2.3mm} 5.971\cdot 10^{-3}\)\sign{***}& \(0.999\)\\
Big-step resizing with warm start & \(  3.186\cdot 10^{4}\)\sign{***}& \( -89.32\)\sign{***} & \( 13.42\)\sign{***} & \(\hspace{2.3mm} 3.202\cdot 10^{-3}\)\sign{***}& \(0.996\)\\
Big-step resizing without warm start  &\( 1.208\cdot 10^{4}\)\sign{***} & \(\hspace{2.3mm} 3.729\cdot 10^{2}\)\sign{***} & \( 18.24\)\sign{***} & \(\hspace{2.3mm} 4.921\cdot 10^{-3}\)\sign{***}& \(0.999\)\\
Constraint changes with warm start & \( 1.074\cdot 10^{5}\)\sign{***}& \( -1.289\cdot 10^{3}\)\sign{***} & \(21.88\)\sign{***} & \(-2.721\cdot 10^{-4}\)\sign{***}& \(0.976\)\\
Constraint changes without warm start      &\(1.231\cdot 10^{5}\)\sign{***} & \( -1.648\cdot 10^{3}\)\sign{***} & \(30.17\)\sign{***} & \( -2.837\cdot 10^{-4}\)\sign{***}& \(0.962\)\\
\bottomrule
\multicolumn{6}{l}{Significance codes: \sign{***} \(p<0.001\)} \\
\end{tabular}
\end{center}
\end{table*}

The variance of the measured runtime differs noticeably for both approaches.
It is smaller for the rather small changes in small-step resizing, and bigger for the big-step resizing and constraint change use cases.
Especially for constraint change, this indicates that some problems with a lot of conflicts were generated, which require more iterations and hence a longer runtime.
The measurements indicate that the variance of the runtime with warm start is smaller than without warm start, and this might be worth further analysis.
It is somewhat astonishing that the experiments reveal only a relatively small effect of the warm-start strategy.
One reason could be the use of the random pivot selector.
Since it selects pivot elements randomly, it can select pivot elements which let the solution deviate strongly from the initial solution before it actually converges towards the new solution.
Another reason can be that the changes in the specification -- even though they are fairly small -- drastically change the solution in some cases.
This can be, for example, due to conflict resolution.
Some constraints, which were not satisfied with the old specification, can become satisfiable and suddenly have an effect on the solution after the specification was changed.
Similarly, small changes in the specification can lead to new conflicts and hence disabling of constraints.

The effects of the warm-start strategy are comparable for all three use cases, but are the strongest for small-step resizing.
This is convenient, as speed is of particular importance for the small-step resizing use case.
Small-step resizing is typically done interactively by the user, and for a good user experience the GUI should react to such resizing in real-time.

\section{Conclusion} \label{sec:conclusion}

In constraint-based GUIs with dynamic behavior, the specification that represents the layout of the GUI is often changed, e.g.\ when a window is resized.
These changes are usually small, resulting in specifications that are very similar.
Since the specifications are similar, one can expect also the results to be similar.
Therefore, we evaluated the use of a warm-start strategy to improve the efficiency of SOR-based constraint solvers for GUIs.
Three common use cases were evaluated with randomly generated GUI layouts:
small-step resizing, big-step resizing, and random changes of several constraints.

We found that an SOR-based solver with a warm-start strategy indeed exhibits a better runtime behavior than a solver without warm-start strategy.
Implementing a warm-start strategy in such solvers does not introduce additional computational effort, as existing values are simply reused.
It is therefore advisable to equip SOR-based GUI layout solvers with a warm-start strategy.

However, we also found that the effect of a warm-start strategy is weaker than we expected.
Possible reasons are the random pivot assignment and the constraint insertion conflict resolution strategy used in the experiment.
Thus, a future work would be to explore the effect of warm starts also for other types of pivot selectors and conflict resolution strategies.
\newpage
\bibliographystyle{IEEEtran}
\bibliography{bibliography}

\end{document}